# COMMENT ON "DISCRETENESS EFFECTS IN SIMULATIONS OF HOT/WARM DARK MATTER" by J. Wang & S.D.M. White

Adrian L. Melott, Dept. of Physics & Astronomy, University of Kansas; melott@ku.edu

ABSTRACT

Wang and White (2007) have discussed some problems with N-body simulation methods. These problems are a special case of a more general problem which has been largely unacknowledged for approximately 25 years, and affects results of all dark matter simulations on small scales (the definition of "small" varying with time).
They present results of a hybrid Tree-PM N-body simulation of hot/warm dark matter-dominated universes, which should have essentially zero initial fluctuation power on a fairly large free-streaming scale dependent upon the dark matter candidate. They analyze the spurious fragmentation of structures on smaller scales comparable to the mean commoving interparticle separation in the simulation. They conclude that such simulations are inaccurate on or below the mean interparticle separation, for both lattice or glass initial conditions.
I emphasize that this result is not restricted to such dark matter candidate models. The mass discreteness limitation has long been described in application to these models, as well as for models with initial power on small scales such as Cold Dark Matter (CDM). Extensive numerical experiments with multiple types of N-body codes have demonstrated that spurious fluctuations due to particle discreteness grow rapidly even in the presence of substantial small-scale power from the intended model spectrum, and modify the results on scales smaller than the mean comoving interparticle separation. This implies that the spatial resolution of such simulations is typically limited not by the force softening length, often referred to as the "resolution", and not by the particle density in halos. Instead it is approximately $N^{-1/3}$, where N is the *mean* particle density, and of course depends on and improves very slowly with increased number of particles.
A partial solution can be constructed by using "nested boxes" of increasing particle density, but this solution is limited by the fact that a typical galaxy is formed from mass scavenged over many $Mpc^3$, and of course a much larger volume for clusters. Reliable N-body simulation results may be achieved by using this strategy, by applying large amounts of computer power, and by a willingness to restrict one's claims to the indicated reliable scales. This calls into question many results on smaller scales over more than two decades.

INTRODUCTION

Wang and White (2007; hereafter WW) simulated Hot Dark Matter (HDM) and several idealized collapse cases in order to study discreteness effects, and to learn whether or not so-called "glass" initial conditions (White 1996) reduce the spurious fragmentation of structures seen earlier. This question is of considerable interest for the current simulation of WMD (Warm Dark Matter), which has damping on a smaller scale than HDM. They concluded that conversion from lattice to glass does not prevent the problem, which is apparent in the formation of spurious clumps on scales comparable to the mean interparticle separation in the simulation. We agree with these conclusions, but argue that except for the extension to the glass initial state, they are a special case of results presented previously, which have much broader implications, including the simulation of CDM and other scenarios with initial small-scale power.

SUMMARY AND COMPARISON OF CONCLUSIONS

This is primarily a review and summary. For detailed information the reader should consult the original literature. The results presented in the summary are more general, in that they include many kinds of tests with many kinds of codes, but they typically do not include the particular glass set of initial conditions which were the focus of WW. There is, however, a general feature of them, the seldom acknowledged limitation due to mass resolution, of which WW is a particular special case.

*(1) Representation of initial conditions*
The sampling of the initial perturbations is carried by the simulation particles. It is naturally subject to the Nyquist sampling theorem. No initial perturbations on smaller scales can be present, although the transition may be sudden or more extended depending upon the unperturbed state. For example, in the Millenium Simulation (Springel et al. 2005) no initial perturbations representing the desired spectrum can exist at wavelengths smaller than about 462 $h^{-1}$ kpc. On these smaller scales, the initial conditions sample the particle discreteness, as emphasized in Splinter et al. (1998, hereafter SMSS), Fig. 5, and WW.

This has not traditionally been reported as part of the initial conditions in papers on N-body simulations, though the final autocorrelation function is often presented on scales *much smaller* than the Nyquist wavelength. The autocorrelation is formally equivalent to the Fourier transform of the power spectrum. Discreteness effects are not readily apparent in such autocorrelations; nonlinear mode coupling typically hides them. This means that they may only be apparent in statistical measures sensitive to phase information. WW were able to show them in the phase-insensitive power spectrum because their cases studied lacked initial small-scale perturbations except from discreteness. The only way to suppress this discreteness and impress the desired initial spectrum instead is to add more particles. Since mode coupling in gravitational instability is much stronger from large to small scales (however see Shandarin & Melott 1990), effects of discreteness are usually large only on scales below the mean interparticle separation.

Knebe et al. (2000; hereafter KKGK) comment that this viewpoint represents a "misconception" since the goal is to follow scales which are resolved in the initial conditions, down to smaller scales as they collapse, by following them with higher force resolution. Since the volume element contracts, the mass resolution inside it is presumed to improve. Unfortunately, as shown by Kuhlman et al. (1996), most volume elements collapse first along one axis; for about of them 90% of them the longest axis has actually *grown (in comoving coordinates)* at the time of first collapse. This is a perfect setup for the kind of spurious fragmentation seen in Melott (1990) and WW. As a result of this scattering, the evolution on small scales is not a consequence of the evolution of the impressed perturbation spectrum.

*(2) Relaxation and other discreteness effects in dynamical evolution*
    (a) *Relaxation*

Efstathiou & Eastwood (1981) performed numerical tests using the $P^3M$ code (Hockney & Eastwood 1981), which like most N-body codes in use today, had force resolution exceeding its mass resolution, in the following sense: The force softening lengthscale $\varepsilon$, on which the interparticle force begins to be reduced below $R^{-2}$, was much less than the mean interparticle separation. They performed tests for relaxation, by running ensembles with particles of varying mass. Relaxation was found to be strong on scales smaller than where the two-point correlation $\xi$ exceeded amplitude of 10; this was close to the mean interparticle separation lengthscale in those simulations.

Suppression of relaxation is extremely important for the faithful modeling of collisionless dark matter. Physical relaxation rates fall to zero as the mass of the component particles drop; N-body simulations of necessity have particle masses far above the values of any dark matter candidate. For example, the typical particle mass in the Millenium Simulation (Springel et al. 2005) is of order $10^9$ Solar masses. Melott (1990) showed the spurious fragmentation of HDM simulations which occurs when the force resolution is smaller than the mean interparticle separation.

PM codes (Hockney & Eastwood 1981) more naturally keep relaxation at a minimum, by sacrificing force resolution so that it is kept comparable (in the above sense) to mass resolution. For this reason the author adopted PM. It enforces a limitation on the extent to which small scales (galaxy formation!) can be resolved simultaneously with large-scale structure. In the spirit of Wittgenstein, "Whereof one cannot speak, thereof one must keep silent." However, it does offer the advantage of making possible much higher particle density, which allowed sampling of large-scale structure, and the first realization that CDM power spectra produce the kind of interconnected structure now called "the Cosmic Web" (Melott et al. 1983; Fry and Melott 1985). The mass segregation test of Efstathiou and Eastwood (1981) was repeated, explicitly applied to PM and published in Peebles et al. (1989). The PM code was run with a "sparse particle" count and a normal one, differing by a factor of 4 in mean interparticle separation. The results of Efstathiou and Eastwood (1981) were confirmed: there was considerable segregation, implying relaxation, on all scales below the mean interparticle separation. Of course, for the "normal" PM run, this was close to the force softening scale, so results would not normally be reported below this scale anyway. But in $P^3M$ and in all other N-body methods in general use, it is common to report results far below the mean interparticle separation. Demonstrably, these results are affected by relaxation and two-body scattering process in general (see also Binney 2004), which are unphysical for nearly all hypothesized dark matter candidates. Since the dark matter dominates the gravitational potential, this problem is transferred to the baryonic component. Of course, relaxation is not the only discreteness problem.

(b) *Plane-wave tests*

Efstathiou et al. (1985) conducted a variety of tests of N-body techniques, emphasizing the improvement in performance with the high force resolution of the $P^3M$ code they used. One of these tests was a plane wave evolution, showing the much better resolution achievable compared with PM. This test was oriented along a coordinate axis, a rather special case.

Melott et al. (1997) extended this to the nonlinear regime, and most importantly followed the collapse of plane-waves which were *not* oriented parallel to one of the three coordinate axes, nor along a cube diagonal, but instead with wavenumbers 2:3:5. This included PM, $P^3M$, and a nested-grid code (adaptive refinement) developed by Splinter (1996). Note that the last code included a high particle density in subgrids, and the mass resolution was kept comparable to the force resolution on these subgrids.

The results were dramatic. The problem is one-dimensional, but did not remain so in general. All runs which had force resolution exceeding mass resolution (Tree, $P^3M$ with $\varepsilon<0.5$ in units of the mean interparticle separation, or PM with sparse particle population) exhibited dramatic unphysical clumping within the sheet. This was of course accompanied by isotropization of what should have been strictly one-dimensional velocity dispersion within the sheet population. In normal PM, in both Tree and $P^3M$ with the force strongly softened, or the nested grid code, these effects were essentially absent. The nested-grid code did the best job of reproducing the total phase space, which was compared with a purely one-dimensional computational nonlinear standard. $P^3M$ with $\varepsilon=0.5$ (a very large value compared with those in general use) showed what might be construed as some improvement: better performance along the test axis with only moderate unphysical scattering. Smaller values of $\varepsilon$ were much worse.

Heitman et al. (2005) also examined plane-wave collapse, but chose a wave aligned along a diagonal of the cube, which has much more symmetry than that of the Melott et al. (2007) case. They found serious problems in replicating the known solution; they argued that the problem was not direct collisionality, but did not examine off-axis motions which would have revealed this.

This is an idealized case, but it is not irrelevant to generalized gravitational instability. As shown by Shandarin et al. (1995) and Kuhlman et al. (1996), gravitational collapse nearly always takes place first in an essentially one-dimensional manner. Therefore, such scattering will be expected *in general, when the force resolution exceeds the mass resolution*, at the very beginning of the nonlinear regime.

(c) *Density and phase cross-correlations*

Splinter et al. (1998; hereafter SMSS) conducted a cross-code comparison between Tree, $P^3M$, and PM methods, with a variety of statistical measures, for a power-law initial perturbation spectrum $P(k) \alpha\ k^{-1}$. This has more power on small scales than CDM, and thus should be expected to set a lower bound to the importance of discreteness effects in such models. It is beyond the scope of this review to describe all the results there. However, a summary of the most important points will be presented. It is important to note that both the mass and the force resolution were varied. That is, the number of particles was varied with $\varepsilon$ adjusted so that the force resolution was on the same absolute scale relative to the simulation volume. Also, $\varepsilon$ was varied while N, the number of particles was held constant.  Identical initial conditions were evolved and then compared between 9 different simulation boxes at 3 different nonlinear stages.

SMSS concluded that although power spectra/autocorrelations are measurably affected by mass resolution, the effects are small and mostly sensitive to force resolution, a conclusion confirmed by Hamana et al (2002).  In particular, they found that when a given code was used, results with varying force resolution (higher than the mass resolution) tended to agree with one another in a cross-correlation test.  KKGK repeated this test and emphasized the internal agreement within most code types.  They noted a disagreement between two versions of PM; however they failed to point out that this disagreement was on scales below the resolution scale anyone would ever claim for a PM code.  On the native PM mesh scale, their cross-correlation amplitude between the two versions was 0.98.  The differences they did find by varying force resolution and code type, to use their words, "applies only to the phase-sensitive statistics".  This is like saying that a certain debilitating disease "only affects right-handed people".  Unfortunately, phases are all-important (e.g. Ma & Fry 2000).  Distributions with the same power spectrum and different phases can be so different as to be unrecognizable (Coles and Chiang 2000).  The only statistics which are *not phase-sensitive* are those that are constructed using only the Fourier amplitudes.  Examples include the power spectrum and its Fourier transform the two-point correlation function, but exclude anything that describes the pattern morphology, such as the bispectrum.   Also, KKGK chose not to vary the mass resolution independently, as did SMSS.

When SMSS varied the mass resolution independently, they found major disagreement of a code with itself until the mass resolution was comparable to the force resolution.  After reaching this limit, codes agreed rather well, even across type.  This agreement was tested by density crosscorrelation and by examining <cos θ>, the mean cosine of the phase angle between Fourier components.  This decreases with increasing wavenumber, and for wavenumbers at the Nyquist limit, this value was around 0.5 for most pairs of runs with high mass resolution, regardless of code type.  For low mass resolution, i.e. softening lengths much smaller than the mean interparticle separation, it was typically about 0.2 *unless* restricted to comparison within a code type.  KKGK interpret this as different kinds of time integration errors, and consequently difficult to eliminate.  Since they did not vary mass resolution they could not detect trends with it nor the convergence evident in SMSS *across* code type, as the mass resolution was increased.

Heitmann et al. (2005) conducted some comparisons as well.  On some statistics, which include effects from longer waves, such as velocity dispersions, they found very good agreement.  On others such as the halo mass function, they found disagreements of 40%.  Disagreements became substantial depending on the codes for halos comprised of 10-100 particles, the lower end corresponding approximately to the mass contained within a Nyquist wavelength as sampled by the particle distribution.

More recently, other tests have found convergence on the mean interparticle separation scale, naturally scaling as $N^{-1/3}$.  As they note, this leads naturally to the use of nested-grid methods in order to resolve smaller scales (Splinter 1996; Diemand et al. 2004; Reed et al. 2005, and references therein).   Such methods do have an inherent limitation, as mass for a given object is scavenged from a large region, which must be fully mass-resolved. But it certainly constitutes a considerable improvement (e.g. O'Shea & Norman 2006). Now that production quality codes allow the mass resolution scale to reach (with certain types of code and massive computer efforts) well inside galaxies, the effects of mass resolution are being noted.

CONCLUSIONS

Elementary considerations of gravitational physics tell us that neither the initial conditions nor the gravitational evolution of simulations with large rest mass can be entirely correct. If, as has been demonstrated many times, there are substantial errors which affect the phases of Fourier components on scales at and below the commoving mean interparticle separation, then there is no reason to trust any new results for which convergence has not been explicitly demonstrated.

ACKNOWLEDGMENTS
I am grateful for helpful comments on a draft of this manuscript by Peter Coles, Hume Feldman, Mikhail Medvedev, and Sergei Shandarin.